\documentclass[12pt,preprint]{aastex}

\shorttitle{Satellite Accretion and the Galactic Halo}
\shortauthors{Brook et~al.}

\begin{document}

\title{Galactic Halo Stars in Phase Space :A Hint of Satellite Accretion?}

\author{Chris B. Brook\altaffilmark{1}, Daisuke Kawata\altaffilmark{1}, Brad K. Gibson\altaffilmark{1} and
Chris Flynn\altaffilmark{1,2}}

\altaffiltext{1}{Center for Astrophysics \& Supercomputing, Swinburne
University, Mail \#31, P.O. Box 218, Hawthorn, Victoria, 3122, Australia}

\altaffiltext{2}{Permanent address: Tuorla Observatory, Piikki\"{o},
FIN-21500, Finland}

\begin{abstract}
The present day chemical and dynamical properties of the Milky Way bear
the imprint of the Galaxy's formation and evolutionary history. One of the most
enduring and critical debates surrounding Galactic evolution is that
regarding the competition between 
``satellite accretion'' and ``monolithic collapse''; the
apparent strong correlation between orbital 
eccentricity and metallicity of halo
stars was originally used as supporting evidence for the latter.
While modern-day unbiased samples no longer support the claims for a
significant correlation, recent evidence has been presented by
Chiba \& Beers (2000,AJ,119,2843) for the existence of a minor population
of high-eccentricity metal-deficient halo stars.  It has been suggested
that these stars represent the signature of a rapid (if minor) collapse
phase in the Galaxy's history.  
Employing velocity- and integrals of motion-phase
space projections of these stars, coupled with a series of N-body/Smoothed
Particle Hydrodynamic (SPH) chemodynamical simulations, 
we suggest an alternative mechanism for
creating such stars may be the recent accretion of a polar orbit dwarf galaxy.
\end{abstract}
\keywords{galaxies: formation --- galaxies: evolution --- Galaxy: halo}

\section{Introduction}

The ``monolithic collapse'' versus ``satellite accretion'' debate
surrounding galaxy formation is a classic one. The former
was best expressed by Eggen, Lynden-Bell \& Sandage (1962, ELS); supporting evidence for the ELS picture came from the
apparent correlation between eccentricity ($\varepsilon$) and metallicity of
halo stars. In contrast, by using a large and reliable set of data for
halo objects chosen without kinematic bias, Chiba \& Beers
(2000, CB00) found no strong correlation between $\varepsilon$ and
metallicity. Bekki \& Chiba (2000) showed that this lack
of correlation is naturally explained by hierarchical clustering
scenarios of galaxy formation. CB00 did, however, identify a concentration
of stars of $\varepsilon$ $\sim 0.9$ and [Fe/H] $\sim -1.7$, and suggested that
this population of stars formed from infalling gas during an early stage
of Galaxy formation, in a manner similar to an ELS collapse.

Current cosmological theories of structure formation have more in common
with accretion-style scenarios, such as that envisioned by Searle \& Zinn
(1978).  
Evidence in support of satellite accretion during our Galaxy's
formation can be found in the observations of considerable substructure in
the halo (e.g. Ibata, Gilmore \& Irwin 1994; Majewski, Munn \&
Hawley 1996; Helmi et~al. 1999; CB00). Helmi \& White (1999, HW99) 
noted that satellites disrupted several billions years ago will 
not retain the
spatial correlations required to allow identification.  However, they
showed that the trail of stars left by a disrupted dwarf satellite will
retain strong correlations in velocity space. Further, the integrals of
motion (E,L,L$_z$) are conserved quantities, and should evolve only
slightly within the relatively stable potential of a galaxy after its halo
is virialized. This space of adiabatic invariants is a natural space to
search for signatures of an accretion event (e.g. Helmi \& de~Zeeuw 2000, HZ00).

We were motivated to run a grid of chemodynamical simulations with the
intention of contrasting the effects of the two collapse scenarios on the
evolution of the Milky Way.  The two models described here vary primarily
in their degree of clustering, and we examine the properties of the
resulting simulated galaxies, in order to uncover present-day
``signatures'' of the models' initial conditions and evolution.

Details of our code and the models used in this study are
provided in Section~2. Section~3 presents the evolution and clustering
histories of the simulated galaxies. We focus on the $\varepsilon$
distribution of our simulated galaxies' halo stars, and on the
comparison to observations. Tracing the evolution of stars originating in
recently accreted satellite galaxies provides a clue to the source of
high-$\varepsilon$ halo stars. We next show that such stars occupy
restricted regions of phase space, a signature of common ancestry. Similar
analysis of the CB00 dataset shows that the identified ``clump'' of
high-$\varepsilon$, low-metallicity, halo stars occupies a similar restricted
region of phase space. The implications for theories of galaxy formation
and their observational signatures are discussed in Section~4.

\section{The Code and Model}

Our Galactic Chemodynamical code ({\tt GCD+}) models self-consistently
the effects of gravity, gas dynamics, radiative cooling, and star
formation. Type~Ia and Type~II supernova feedback is included, and the instantaneous chemical recycling approximation is relaxed. Details of {\tt GCD+} can be found in Kawata \& Gibson
(2003).
  
The semi-cosmological version of {\tt GCD+} used here is based upon the
galaxy formation model of Katz \& Gunn (1991).  The initial condition is
an isolated sphere of dark matter and gas, onto which small scale density
fluctuations are superimposed, parameterized by $\sigma_8$.  These
perturbations are the seeds for local collapse and subsequent star
formation. Solid-body rotation corresponding to a spin parameter
$\lambda$ is imparted to the initial sphere. For the flat CDM model described
here, the relevant parameters include $\Omega_0 = 1$, $H_0 = 50$km s$^{-1}$ Mpc$^{-1}$, total mass ($5\times
10^{11}$~M$_\odot$), baryon fraction ($\Omega_{b}=0.1$), and spin
parameter ($\lambda=0.0675$). We employed 38911 dark matter and 38911
gas/star particles, making the resolution of this study comparable to 
other recent 
studies of disk galaxy formation (e.g. Abadi et~al. 2003).  The two
models described here differ only in the value of $\sigma_8$. In model~1 (M1),
$\sigma_8=0.5$, as favoured in standard CDM ($\Omega_0 = 1$) cosmology. In
model~2 (M2), we explore the use of a smaller value for
$\sigma_8$ of 0.04, a value which results in a more rapid,
dissipative, collapse.

\section{Results} 

Figure~1 illustrates the evolutionary histories of the two models. M1 demonstrates
classical hierarchical clustering - gas collapses into the local dense regions
seeded by the initial small-scale perturbation, with star formation occurring
subsequently in
these over-dense regions.  Stars continue to form in sub-clumps, as
well as in the central region as a disk galaxy is built up.  We see less
clumping in M2, with gas streaming homogeneously toward the centre of
the galactic potential, resulting in most of the star formation occurring
in the central regions. 

We analysed the bulk properties of our simulation at $z = 0$ and confirmed
that they were consistent with the simulations of Katz (1992), Steinmetz
\& Muller (1995), Berczik (1999), and Bekki \& Chiba (2001).  The predicted
surface density profiles, metallicity gradients, specific angular momenta of gas and stars, and rotation curves for
our models did not differ significantly from previous studies, or between our
own two
models. We suffer from the same overcooling problem encountered in
earlier studies (White \& Frenk 1991). This results in too high rates of
star production at early epochs, overly rapid metal enrichment, and a halo
metallicity distribution function peaked approximately one dex higher than
that observed (Ryan \& Norris 1991).  The field of MDFs is the subject of a future, lengthier, analysis,
but we do note that this offset in the halo MDF peak does not impact on
our analysis here (we can still adopt a differential metallicity cut
to delineate between the halo and thin disk in our simulation).
 We did  find a difference in the orbital $\varepsilon$
distribution of the simulated solar neighbourhood halo stars of the two models.
After Bekki \& Chiba (2000), individual stellar eccentricities were calculated
by allowing the orbits of star particles to evolve
for $1.8$~Gyrs under the gravitational potential of the simulated disk
galaxy achieved at $z = 0$.  Eccentricities 
were then derived using
$\varepsilon=(R_{apo}-R_{peri})/(R_{apo}+R_{peri})$, where $R_{apo}$ 
($R_{peri}$) is the apogalactic (perigalactic) distance from the
centre of the simulated galaxy.  Halo stars are defined to be simulated star particles with [Fe/H] $<
-0.6$. Due to the higher-than-expected absolute value for the halo MDF
(alluded to earlier), this metallicity cut may lead to an underestimate
of the halo population, but the final halo "sample" is representative.

The dotted (dashed) line of
the histogram in Figure~2a shows the $\varepsilon$ distribution of halo
star particles in the solar circle of M1 (M2).
The solar circle is defined as an annulus of the galactic disk\footnote{The volume of our  "solar circle" is roughly an order-of-magnitude larger than the analog employed in the CB00 dataset.  We have ensured that this apparent volume "mismatch" does not impact upon our results by testing for sensitivity to azimuthal variations within the annulus by subdividing the simulated dataset into random octiles. The  phase-space effects described are robust to this azimuthal subdividing.}, bounded by
$5<$ R$_{XY}<12$ kpc and $|Z| <2$ kpc, where $Z$ is the rotation
axis and R$_{XY}=\sqrt(X^2+Y^2)$. Each bin shows the fraction of such star
particles lying in a given $\varepsilon$ range.  Also shown (solid line)
is the observational constraint from CB00.  M1 produced a greater number of
high-$\varepsilon$ ($\varepsilon$ $>$ 0.8) solar circle halo stars, and is in
better agreement with observation. 

Considering the analysis of ELS, whose rapid collapse model resulted
in highly eccentric halo stars, it was not unreasonable to expect that the
more dissipative, ``monolithic'' nature of the collapse of M2 would
result in a higher number of high-$\varepsilon$ halo stars. This prompted a
more detailed examination of the collapse time of M2. Using analysis
of the evolution of top hat overdensities with the collapse redshift $z_c
= 1.75$, which we adopt (Padmanabhan 1993), the collapse timescale
$t_{coll}$ of M2 is estimated to be $\sim$1.4 Gyrs, thus not satisfying the conditions originally proposed by ELS, $t_{coll} <<
t_{sf}$ (where $t_{sf}$ is the star formation timescale). While encompasing the spirit of ELS within the current cosmological paradigm, M2 should not be considered an exact analogue of the ELS rapid collapse scenario.

As the difference between models~1 and
2 is in the clustering history, our result suggests that the Milky Way
may have experienced significant clustering processes during its formation.
To clarify this hypothesis, we examine the specific accretion history of
each model, tracing the $\varepsilon$ distribution functions for the stars
associated with each disrupted satellite. In M1 we identify
two satellites at $z = 0.46$ which have merged into the halo of the host
galaxy by $z = 0$. These are the final significant merger events in the simulated galaxy's formation. 
The 
largest such dwarf galaxy (hereafter S1), with stellar mass $\sim 10^9$~M$_\odot$ , was on 
on a precessing polar orbit\footnote{The stellar
mass of the satellite appears too large considering the inferred mass of
the Galaxy's stellar halo (e.g. Morrison 1993).  However, the {\it fractional} contribution of S1 to the simulated halo ($\sim$10\%)
is reasonable, even if the {\it absolute} mass is overestimated.  This
overestimation is an artifact of the overcooling problem alluded to
earlier, and does not impact upon the phase-space conclusions described
here.}.  By $z = 0.26$, tidal forces had begun
stripping stars from S1, and by
$z = 0$, its stars were spread effectively 
throughout the halo. 

The histogram of
Figure~2b shows the $\varepsilon$ distribution of solar circle halo stars
which originated in the satellites identified at $z = 0.46$. The
$y$-axis is normalised by the total number of solar circle halo stars in
each $\varepsilon$ bin. The dashed line is those stars from S1. We notice that the majority
of these halo stars are of high-$\varepsilon$, and that S1 in
particular contributes $\sim$20\% of all high-$\varepsilon$ halo stars in
the solar circle at $z = 0$.

In Figure~3 we plot the phase space distributions of solar circle stars of
M1.  Figures~3a-c show: velocity directed radially away from the
galactic center V$_{rad}$ versus rotational velocity in the plane of the
disk V$_{\phi}$; velocity out of the plane V$_z$ versus V$_{\phi}$; and
V$_{rad}$ versus V$_z$.  Figure~3d plots the integrals of motion, 
projected angular
momentum L$_z$ versus the absolute value of the 
energy ($|$E$|$). Stars originating
in S1 are marked as open
circles. We see that such stars occupy a restricted region of phase space.  
It is worth noting that these results are qualitatively similar to those
of HW99 and HZ00, both of
which follow (at higher spatial resolution) the accretion
of stars in a {\it static} potential, 
in which little disruption to the adiabatic
invariants would be expected. Thus our study confirms that these results
hold when the formation processes of the galaxy are self-consistently
simulated, and hence the potential in which the satellite is disrupted is
dynamic. In fact, our final velocity phase space distribution (Figure~3a-c)
is reminiscent of Figure~5 of HW99, bearing in mind that most of our 
stellar orbits will be at intermediate points between their pericenter and
apocenter. Further, the distribution of the disrupted stars in integrals of
motion
space (Figure~3d) is in good qualitative agreement with those
for satellites in Figure~7 of HZ00.

Motivated by the high-$\varepsilon$ seen in the accreted satellite 
stars (Figure~2b), we examined the phase space distribution of the
clump of high-$\varepsilon$ metal-deficient 
halo stars identified by CB00. Figure~4 shows the sample of CB00 stars
within $2.5$~kpc of the Sun with [Fe/H] $\le -1$, plotted in the same
phase space projections as the simulated data in Figure~3.
Highlighted by open circles are
stars with $\varepsilon$ $> 0.8$ and $-2.0<$ [Fe/H] $<-1.4$. 
A wider range of energies is seen in the observed dataset, in comparison with 
the simulations shown in Figure~3d.  The latter
indicate that recently accreted satellite stars will be of higher
energy (lower $|$E$|$) on average than halo field stars, as their stellar orbits will be
less bound. We thus make an arbitrary cut in energy in order to separate
the field stars from those we suspect come from an accreted satellite, and
the new subset of stars are marked by solid circles. We notice immediately
that these stars are more tightly confined in phase space, and closely
resemble the distribution of the accreted satellite in our simulation.

\section{Discussion}

We have simulated self-consistently two models of
the formation of a Milky
Way-like galaxy using our N-body/SPH chemodynamical code {\tt GCD+}. The models
differ only in the amplitude of small scale density perturbations $\sigma_8$
incorporated into the initial conditions. This results in different
merging histories, which we then use to search for distinctive present-day
chemical and kinematical signatures of the simulated galaxies' early
formation epochs.
In M1, stars form in local dense regions of gas,
seeded by a value of $\sigma_8$ comparable to current CDM orthodoxy. The
galaxy subsequently builds up via a series of mergers in the manner of
standard hierarchical structure formation scenarios. A low value
of $\sigma_8$ in M2 resulted in a more dissipative collapse of the
baryonic matter, with star formation occurring predominantly in the
galaxy's central regions. Both models result in disk galaxies with
remarkably similar qualities; M1, however, produces a greater number
of high-eccentricity halo stars, and in this regard is in better agreement
with the observational dataset of CB00. By tracing the merging histories of
satellites in M1, it becomes apparent that a significant number of
high-$\varepsilon$ halo stars originated in accreted dwarf galaxies. In
fact, almost 20\% of high-$\varepsilon$ halo stars located in the solar
circle originated in one satellite, of stellar mass $\sim$10$^9$~M$_\odot$,
which was on a precessing polar orbit and was accreted over the past
$\sim$5~Gyrs.

Thus our simulation suggests that stars from an accreted satellite which
was on a polar orbit can form part of the galaxy's halo, and that such
stars have highly eccentric orbits. Originating in accreted polar
satellites, these stars would not gain the angular momentum induced by the
tidal torque of the external gravitational field which results in the
rotation of the disk component of the galaxy. Such stars will fall toward
the centre of the galactic potential and end up on highly eccentric
orbits. 

The existence of an apparent
correlation between metallicity and $\varepsilon$ in
halo stars was taken by ELS to be a sign of a rapid collapse driving the
formation of the Galaxy. Our study suggests that another way of creating
such high-$\varepsilon$ halo stars is through the recent accretion of polar
orbit satellites.  Furthermore, dwarf galaxies of stellar mass 
$\sim 10^9$M$_{\odot}$ typically have appropriate average metallicities to be
the source of this group of stars  with [Fe/H] $\sim -1.7$ (e.g. Mateo 1998). Velocity phase space and the space of
adiabatic invariants is where we would expect substructure within the halo
due to past merging to be apparent. The distribution in such space of an
accreted satellite of our simulated galaxy resembles that of previous
studies which traced the disruption of dwarf galaxies in static
potentials. 
The greater dispersion in the
distributions of our simulation can most likely be attributed to the
dynamic nature of the potential in our full simulation of Galactic
evolution, eroding the invariance of the integrals of motion. The
identified concentration of high-$\varepsilon$, low-metallicity, halo stars
from CB00 occupy a similar, restricted, region of phase space to the
simulated accreted satellite.

The suggestion that this concentration of stars in the
$\varepsilon$-metallicity plot is caused by satellite accretion has also
been noted by Dinescu (2002).  In the metallicity range $-2.0 <$
[Fe/H] $< -1.5$ of the CB00 data, Dinescu notes an excess of stars in
retrograde orbits with rotational velocity (V$_{\phi}$) 
$\sim -30$~km~s$^{-1}$,
from that expected from a ``pure'' halo. She identifies these stars as
candidates for having been torn from the system that once contained
$\omega$~Cen. $\omega$~Cen is believed to be the nucleus of a disrupted, accreted dwarf galaxy (e.g. Lee et~al. 1999), is in a retrograde orbit, and has an orbital
$\varepsilon$ of $\sim$0.67.  Using a simple disruption model, Dinescu
showed that the tidal streams tend to have higher orbital $\varepsilon$
than the disrupted satellite. Along with the mean metallicity of 
$\omega$~Cen being [Fe/H] $= -1.6$, the inference is that the clump of
high-$\varepsilon$ halo stars identified in CB00 contains a significant
number of stars which have been tidally stripped from $\omega$~Cen. Our
simulations provide a degree of support for this hypothesis. Taking those
stars with L$_z < 0$ in Figure~4d, we see a clump of stars with
intermediate energies and L$_z \sim -400$~kpc~km~s$^{-1}$. These stars are
identified in all phase space plots by grey circles;  we see that the
distribution becomes more restricted in each velocity space dimension, and
there is greater concorde with HW99's
idealized satellite accretion studies.

The key question we wish to address remains ... \it what are the
implications for the competing galaxy formation paradigms? \rm A brief
response is as follows: CB00 observationally found no correlation between
eccentricity and metallicity for halo stars near the Sun, obviating the
need for a ``rapid collapse'' picture of Galaxy formation. However, CB00 do interpret a clump of high-eccentricity
low-metallicity stars in this observational plane in
terms of ELS - i.e. as a relic of a rapid collapse phase. Our simulations
suggest that an equally plausible origin for this clump is the recent
accretion of a polar orbit satellite in the Galactic halo.
 
\acknowledgments 
The authors thank Masashi Chiba and Tim Beers for providing the data for
Figure~4. We appreciate a discussion with Ken Freeman, and the
helpful suggestions of the anonomous referee, Mike Beasley, and Amina Helmi. This study made use of the
Victorian and Australian Partnerships for Advanced Computing and was supported 
by the Australian Research Council through its Large Research Grant
(A0010517) and Linkage International Award (LX0346832) programs. CBB is
funded by an Australian Postgraduate Award. CF thanks the Academy of
Finland and its ANTARES space research program.

\clearpage

\begin{figure}
\plotone{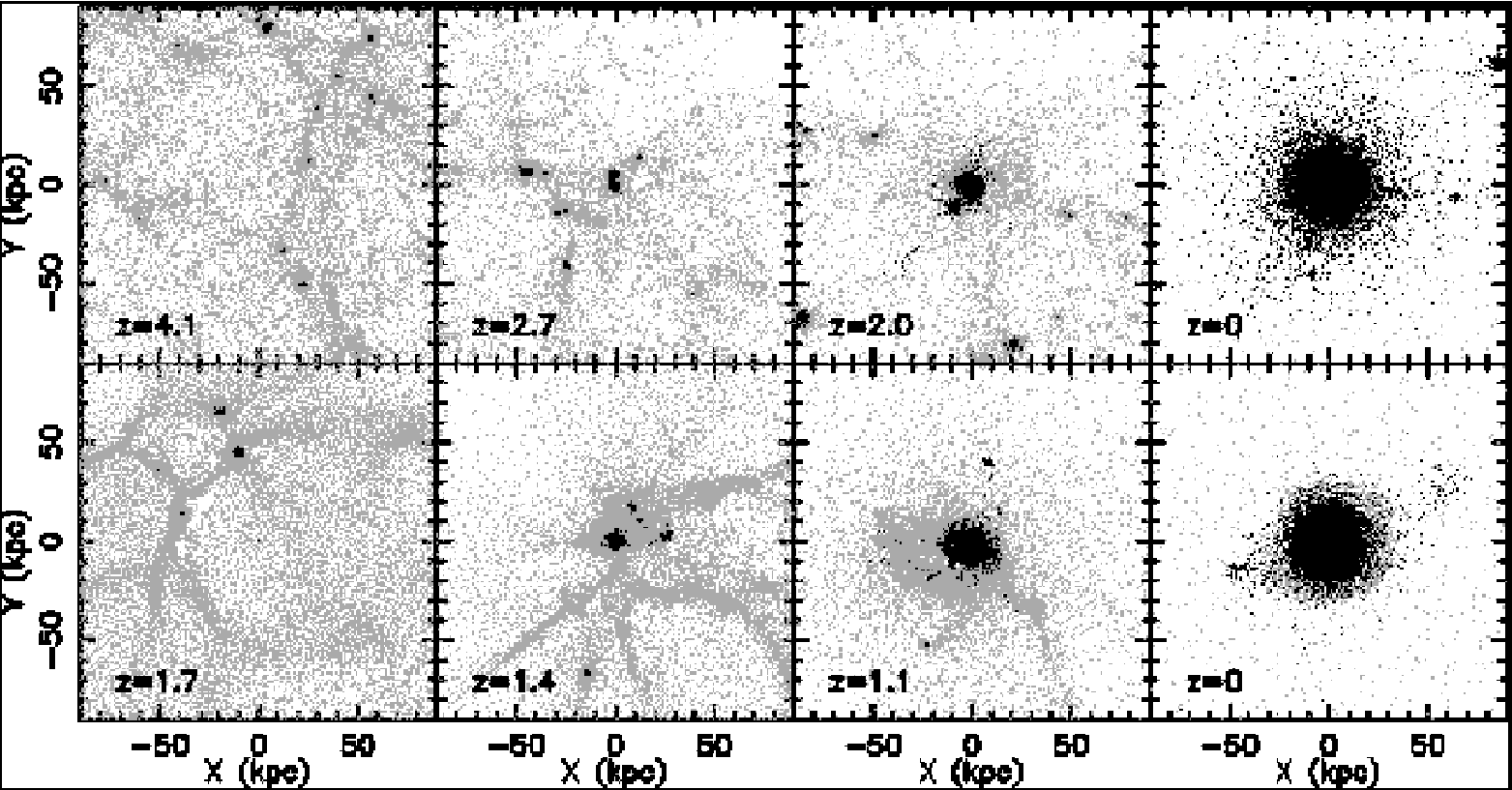}
\caption{$X-Y$ plots of M1 (upper row) and M2.
The $Z$ axis is defined to be the initial axis of rotation. 
Grey (black) dots represent gas (star) particles. Epochs are chosen 
(with redshifts labeled) so that roughly the same stellar mass 
is present in corresponding upper and lower panels.
Gas collapse and star formation are more centralised in M2.  \label{fig1}}
\end{figure}

\clearpage 

\begin{figure}
\plottwo{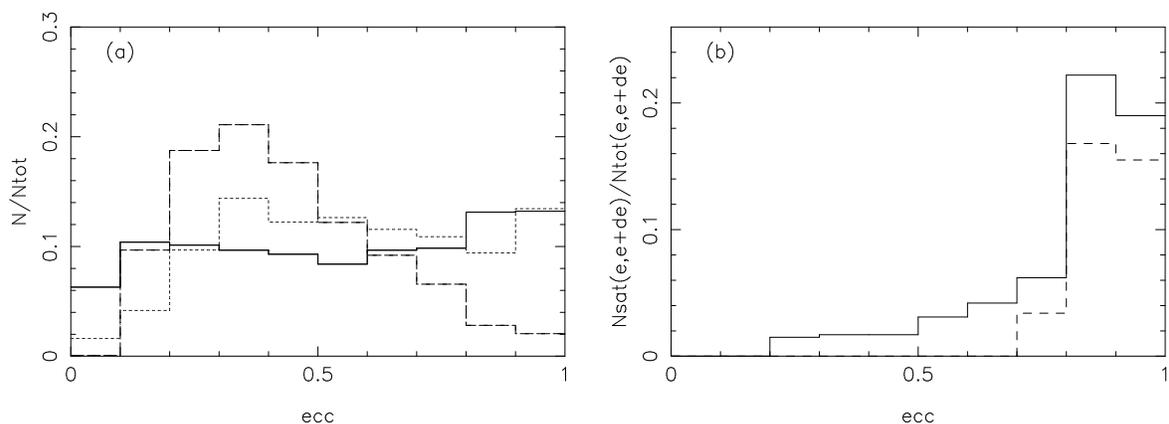}{f2b.ps}
\caption{$(a)$:
Present-day $\varepsilon$ distribution of halo stars at the solar circle for 
our two models - the dotted (dashed) line represents M1 (M2). 
The solid line corresponds to the observational dataset of 
CB00. M1 leads to a greater
number of high-$\varepsilon$ halo stars in the solar circle. $(b)$: The
solid line shows the $\varepsilon$ distribution of solar circle halo stars
which were in satellites at redshift $z = 0.46$. The dashed line shows stars
originating from a single such satellite (S1). The
$y$-axis is normalised by the total number of star particles in each
$\varepsilon$ bin (from Figure~2a).\label{fig2}}
\end{figure}

\clearpage 

\begin{figure}
\plotone{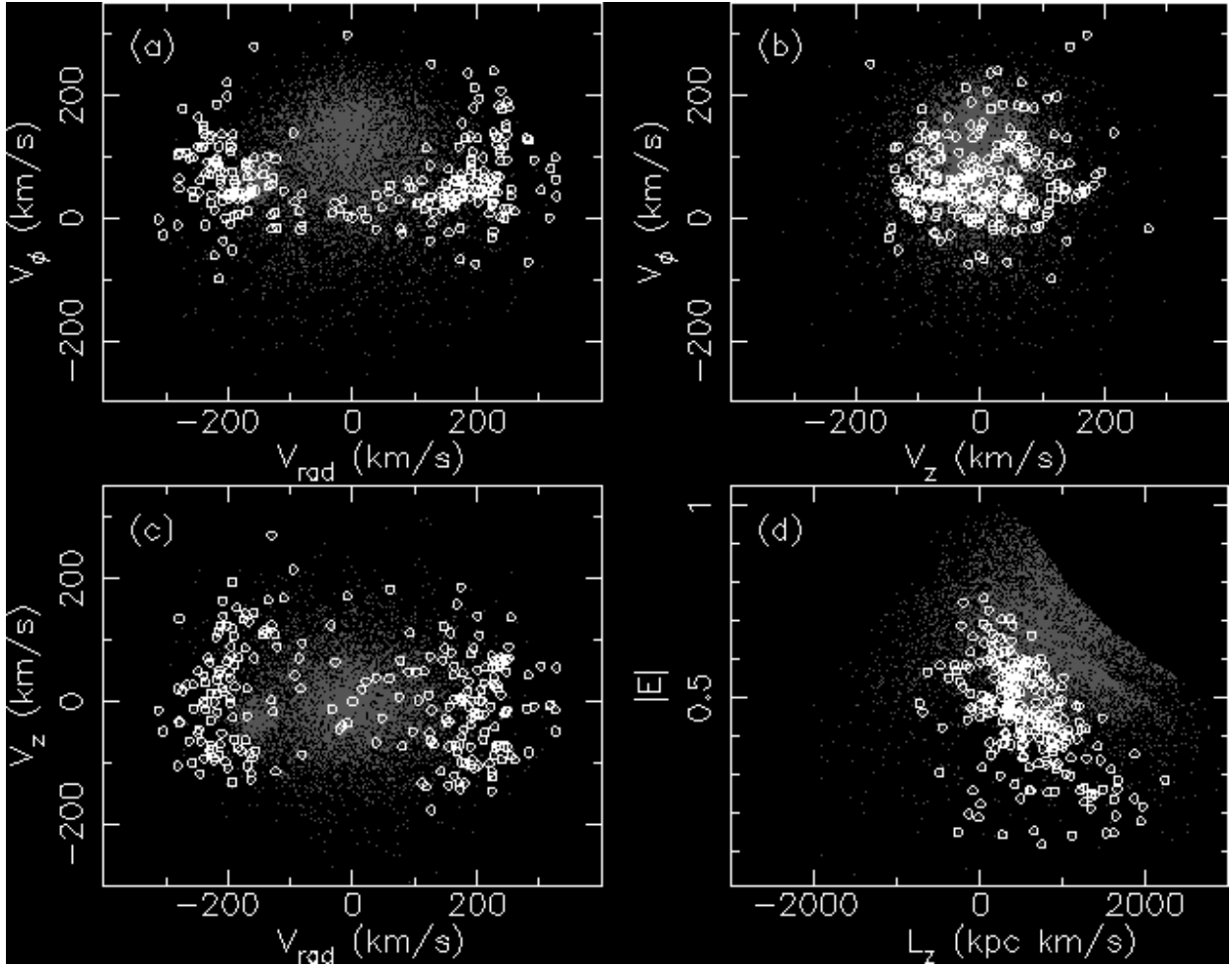}
\caption{$(a)-(c)$: 
Velocity space  projections of the 
present-day solar circle
stars of M1. Open circles denote the subset of stars originating
in S1. $(d)$: The distribution for
the same stars in integrals of motion  space - i.e., the absolute value of the
total energy ($|$E$|$)  
versus projected angular momentum (L$_z$).  \label{fig3}}
\end{figure}

\clearpage

\begin{figure}
\plotone{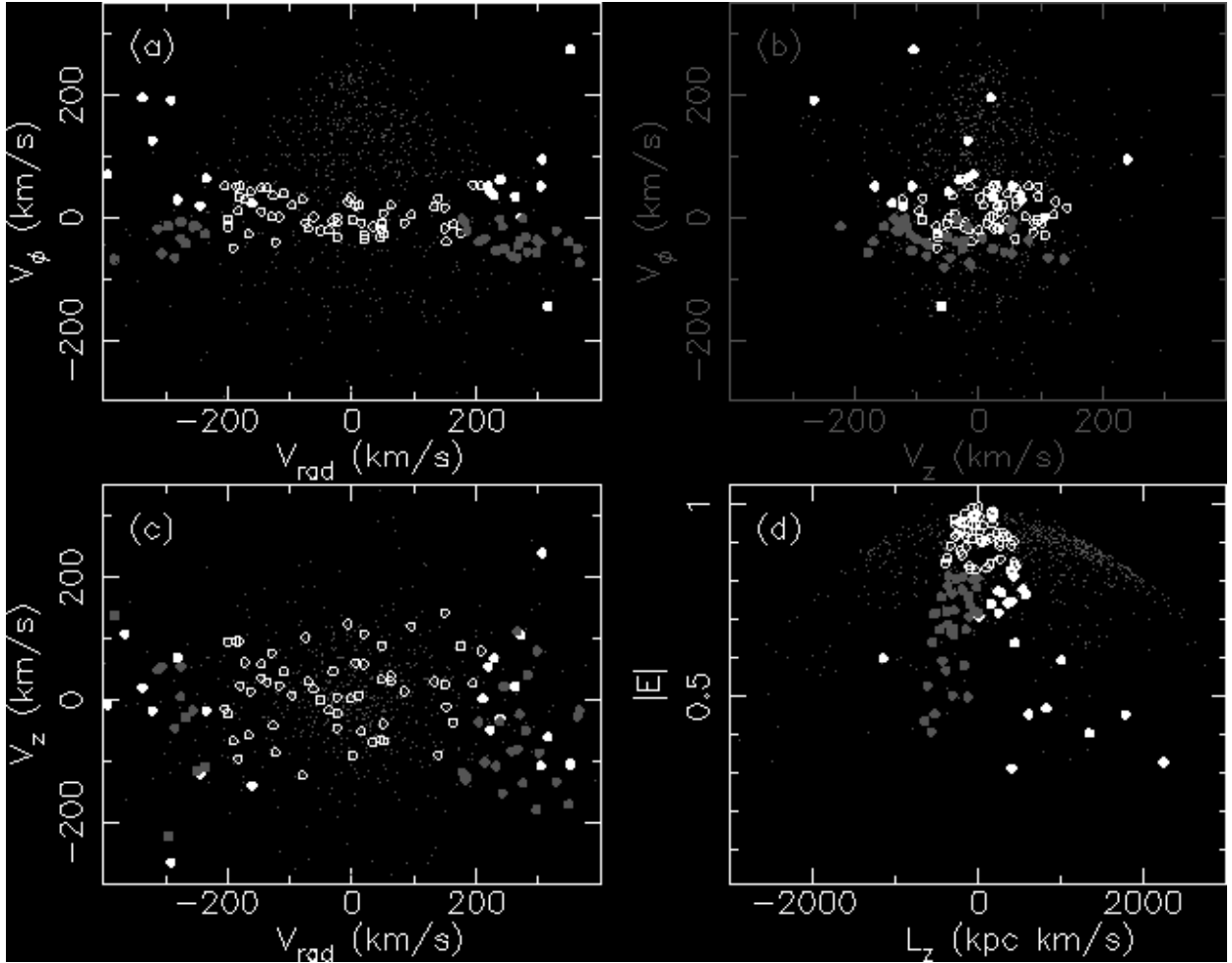}
\caption{Velocity and integrals of motion
phase space for the 729 CB00
stars within $2.5$~kpc of the Sun with 
[Fe/H] $<-1$. $(a)-(c)$: velocity space; $(d)$: energy versus
versus projected angular momentum. Stars with $0.8<$ $\varepsilon$ $<1.0$ and  $-2.0<$ [Fe/H]
$<-1.4$ are represented as open circles. 
Solid  circles denote stars of high
energy; grey circles correspond to high energy stars with 
$-800<$ L$_z$ $<0$~kpc~km~s$^{-1}$.  \label{fig4}}
\end{figure}

\clearpage

\end{document}